\newcommand{\NA}{N_{\rm A}}
\newcommand{\rme}{{\rm e}}
\newcommand{\Cs}{{^{133}\rm Cs}}
\newcommand{\Rb}{{^{87}\rm Rb}}
\newcommand{\rmC}{{^{12}\rm C}}
\newcommand{\rmS}{{^{32}\rm S}}
\newcommand{\Si}{{^{28}\rm Si}}
\newcommand{\rmX}{{\rm X}}
\newcommand{\rmH}{{\rm H}}
\newcommand{\rmD}{{\rm D}}
\newcommand{\rmn}{{\rm n}}
\newcommand{\dd}{d_{220}}

\documentclass[]{amsart}
\usepackage[english]{babel}
\usepackage{amsmath,amsthm}
\usepackage{amsfonts}
\usepackage{graphicx}
\usepackage{epstopdf}

\begin{document}

\title{Precision measurements of the Planck and Avogadro constants}%
\author{H.\ Bettin$^1$, K.\ Fujii$^2$, J.\ Man$^3$, G.\ Mana$^4$, E.\ Massa$^4$, and A.\ Picard$^5$}%
\address[1]{PTB -- Physikalisch-Technische Bundesanstalt, Bundesallee 100, 38116 Braunschweig, Germany}
\address[2]{NMIJ -- National Metrology Institute of Japan, 1-1-1 Umezono, Tsukuba, Ibaraki 305-8563, Japan}
\address[3]{NMI -- National Measurement Institute, Bradfield Road, Lindfield, NSW 2070, Australia}
\address[4]{INRIM -- Istituto Nazionale di Ricerca Metrologica, str.\ delle Cacce 91, 10135 Torino, Italy}
\address[5]{BIPM -- Bureau International des Poids et Mesures, Pavillon de Breteuil, 92312 S`evres Cedex, France}

\begin{abstract}
Precision measurements of the fundamental constants are tour de force of basic metrology, where the useful information is usually beyond the last digit of the measured value. They challenge theoretical models and measurement technologies and set a network of measurement equations on which a universal system of units can be built, which stems from the most basic concepts of physics. Because of their connection with the mass unit, the Avogadro and Planck constants are on the spotlight.
\end{abstract}

\maketitle

\section{Introduction}
Since 1889, the international prototype of the kilogram -- a Pt-Ir artefact -- serves as the definition of the unit of mass of the International System of units (SI): The kilogram definition states that 1 kg is the mass of the international prototype. Therefore, the international-prototype mass is a {\it sui generis} type of fundamental constant, which affects the SI value of all the quantities related to mass and energy. The international prototype is stored in a vault of the Bureau International des Poids et Mesures, with a number of official copies. Its mass was compared to that of its copies and the national prototypes at intervals of about 40 years; the results show divergences with time of an average of 50 $\mu$g over one hundred years \cite{Girard:1994}. We might conclude that the value of all the mass- and energy-related quantities, like the electron mass and the Planck constant, is similarly drifting with time, but more prosaically the data might indicate that the prototype is losing contaminants or platinum or iridium. Consequently, a drift of the SI value of the mass- and energy-related fundamental constants would be only apparent.

Since the measurement uncertainty of the related fundamental constants is reaching the uncertainty of mass measurements, this situation is no longer acceptable. Therefore, it has been planned to redefine the kilogram by fixing the numerical value of the Planck constant, $h$ \cite{Mana:2012}. Fundamental constants -- for example, the speed of light $c$ in the Einstein equation $E=mc^2$ -- are concept synthesizers equating seemingly different quantities. Accordingly, they are conversion factors between measurement units -- in the previous example, the mass and energy units -- and are of the utmost interest in metrology. In 2011, the 24th General Conference on Weights and Measures sanctioned that the International System of units will be upgraded in terms of fundamental constants: Four of the base units -- the kilogram, ampere, kelvin, and mole -- will be redefined in terms of fixed numerical values of the Planck constant, the elementary charge, the Boltzmann constant, and the Avogadro constant. In addition, specific {\it mises en pratique} will be issued to describe how to realize the units in a practical way. For example, any experiment that is used today to measure the SI value of the Planck constant -- hence, necessarily, in terms of the mass of the international prototype -- will be reversed to give the prototype mass in terms of an internationally agreed numerical value of the Planck constant.

From a practical viewpoint, it is necessary that the relative realization-uncertainty -- which is the same as the uncertainty of the $h$ measurement -- does not make the unit dissemination to science and industry worse than what it is today. In addition, it is necessary that the unit redefinition is invisible to most of the users, apart the metrologists themselves. This requires that the accuracy of the $h$ measurements is at least $2\times 10^{-8} h$ and that the $h$ value is chosen in such a way that the mass of the international prototype is indistinguishable, when traced back to the new definition, from 1 kg to within the same uncertainty. To this end, the $h$ value will be set to its best estimate calculated by a least-squares adjustment of the measured values of the fundamental constants. The last adjustment was completed in 2010 \cite{CODATA:2010}; the recommended Planck-constant value is $h=6.62606957(29) \times 10^{-34}$ J s.

The precision measurements and least-squares adjustment of the fundamental constants probes the Nature's working by checking the internal consistency of the network of relationship established by its mathematical description. The energies where the effects of the fundamental interactions are expressed could be out of the reach of any technology, but it may be possible to look for minute effects at accessible energies by measurements of outstanding sensitivity and accuracy. The adjusted constant-values are fitted to input-data obtained from widely differing experiments and by assuming that a number of interpretative models and measurement equations are valid. Therefore, the least-squares adjustment tests the correctness and consistency of these models and of the relevant measurement technologies.

This paper summarizes the methods to measure the ratio between the Planck constant and a mass; it reviews the watt-balance experiment, where the integration of mechanical and electrical measurements allows $h/m(\mathfrak{K})$, $m(\mathfrak{K})$ being the mass of the international prototype of the kilogram, to be determined. Next is a review of the $^{28}$Si experiment, where the Avogadro constant is determined by counting the atoms in a silicon crystal highly enriched by the $^{28}$Si isotope. By rewriting the $h/m$ ratio, where $m$ is the mass of a particle or an atom, as $\NA h/M$, where $M$ is the molar mass and $\NA$ the Avogadro constant, the $h$ values obtained from atomic- and nuclear-physics measurements can be compared with those obtained from the watt-balance experiments, which relies on solid state physics. As many times happened in the past, there is an inconsistency between the $h$ determinations, which indicates that an error was made in at least one of the measurements. Hence, there is room for developments to resolve this discrepancy. The last section looks at the possible weakness in counting the silicon atoms. It illustrates the activities necessary to investigate the critical points of the experiment and to stress the $\NA$ measurement and the relevant know-how with a challenging goal to reduce the uncertainty to $1.5\times 10^{-8}\NA$.

\section{The measurements of the Planck constant}
The measurement of the Planck constant is starting a new phase. Quantum mechanics shows that the Planck constant links the wave-function energy $E$ and frequency $\nu$ by the Planck equation $E=h\nu$. Consequently, $h/c^2$ is the conversion factor between frequency and energy units. Additionally, when quantum mechanics is combined with relativity, $h$ links also the Compton frequency $\nu_C$ of a relativistic matter-wave (in the reference frame where matter is at rest) to its mass-energy $E=mc^2$. Consequently, by combining the Planck and Einstein equations, we obtain $h\nu_C=mc^2$, which associates a frequency to any mass and shows that $h$ is also the conversion factor between frequency and mass units.

The measurements of the $h/m$ ratio can be traced back to frequency or wavelength measurements. In principle, $h/m$ could be measured by annihilating an electron-positron pair and determining the frequencies of the emerging photons. However, this frequency is far beyond today technology; we require a method of scaling it into an accessible range.

A simple example of how this can be done is illustrated by the $h/\Delta m(\Cs)=c^2/\nu(\Cs)$ equation, where $\Delta m(\Cs)$ is the mass defect between the two hyperfine energy levels that define the second and $\nu(\Cs)=9192631770$ Hz by definition. A less simple measurement equation is $h/m(\rme) = c\lambda_C(\rme) = c\alpha^2/(2R_\infty)$, where $\lambda_C(\rme) = c/\nu_C(\rme) = \alpha^2/(2R_\infty)$ is the Compton wavelength of the electron and $\alpha$ and $R_\infty$ are the fine-structure and Rydberg constants. Since both $\alpha$ and $R_\infty$ are extremely well measured, $h/m(\rme)$ can be calculated to with a $6.4 \times 10^{-10}$ relative uncertainty.

Another example is the $h/m(\rmn) = 3.956033285(287) \times 10^{-7}$ m$^2$/s ratio, whose measurement was completed in 1998 \cite{Kruger:1995,Kruger:1998,Kruger:1999}. This experiment relied on the de Broglie equation $m(\rmn)u = h/\lambda$, where both the wavelength $\lambda$ and the velocity $u$ of monochromatic neutrons were measured. Monochromaticity was obtained by Bragg reflection on a calibrated silicon crystal, whereas the neutron velocity was determined by time-of-flight measurements.

As regards measurements based on atom masses, the $h/m(\rmX)$ ratio has been determined by measuring the recoil velocity $u$ of $\Cs$ and $\Rb$ atoms absorbing or emitting photons in Ramsey-Bordé interferometers \cite{Chu:2002,Nez:2008,Nez:2011,Mueller:2013}. The photon momentum $h/\lambda$, where $\lambda$ is the photon wavelength, is balanced by the atom recoil-momentum $m u$. Conservation of momentum yields $h/m = \lambda u$. The interferometer operation can be also viewed as a measurement of the difference between the proper times of the atoms propagating through the interferometer arms, which is equivalent to measure a subharmonic of the Compton-frequency of the associated matter-waves \cite{Mueller:2013}.

Measurements based on nuclear physics are possible by determining the wavelengths of the $\gamma$ photons emitted in the cascades from the capture state to the ground state in the neutron capture reactions \cite{Rainville:2005}
\begin{equation}\label{X}
 \rmn + {^n\rmX} \rightarrow {^{n+1}\rmX^*} \rightarrow {^{n+1}\rmX} + \sum \gamma ,
\end{equation}
where, for instance, $\rmX$ is either $\rmS$ or $\Si$, and
\begin{equation}
 \rmn + \rmH \rightarrow \rmD^* \rightarrow \rmD + \gamma_{\rmD^* \rightarrow \rmD} .
\end{equation}
The $h/m(\rmC)$ determination is based on the facts that the daughter isotope is lighter than its parents and that the mass defect can be measured by determining the wavelengths of the $\gamma$ rays emitted in the decay of the capture state to the ground state. Wavelengths are determined by a double crystal diffractometer in terms of the calibrated lattice parameter of the diffracting crystals. The measurement compares the total energy of the emitted $\gamma$ photons against the mass defect between the capture and ground states. With a few algebra, the comparison can be written as
\begin{equation}
 \frac{h}{m(\rmC)} = \frac{\left[ A_r({^n\rmX}) + A_r(\rmH) - A_r({^{n+1}\rmX}) - A_r(\rmD) \right] c^2}
 {\sum \nu_{^{n+1}\rmX^* \rightarrow ^{n+1}\rmX} - \nu_{\rmD^* \rightarrow \rmD}} ,
\end{equation}
where $A_r$ are the relative atomic mass and $\nu=c/\lambda$ are the $\gamma$-ray frequencies. The measurements of the relative atomic masses are carried out by comparing the cyclotron frequencies of the relevant ions confined in a Penning’s trap \cite{Blaum:2006}.

A direct way of measuring the $h/m(\mathfrak{K})$ ratio, where $m(\mathfrak{K})$ is the mass of a kilogram prototype, is by a watt balance \cite{Robinson:2012,Steiner:2013}. This experiment compares virtually the mechanical and electrical powers produced by the motion of a kilogram prototype in the Earth gravitational field $g$ and by the motion of the supporting coil of length $L$ in a magnetic field $B$. The comparison is carried out in two steps. In the first step, a balance is used to compare the prototype weight $m(\mathfrak{K})g$ with the force $BLI$ generated by the interaction between the electrical current in the coil, $I$, and the magnetic field. The measurement of $I$ is based on the Josephson and quantum Hall effects. Hence, $I = U/R = n_1n_2\nu e/2$, where $U = n_1h\nu/(2e)$, $R = h/(n_2e^2)$, $e$ is the electron charge, $\nu$ is the frequency of the microwave irradiating the Josephson device, and $n_1$ and $n_2$ are integers. In the second step, the coil is moved with velocity $u$ and the induced electromotive force, $\mathcal{E} = BLu$, is measured. The measurement of $\mathcal{E}$ is based again on the Josephson effect, that is, $\mathcal{E} = n_3h\nu/(2e)$, where $n_3$ is an integer. Eventually, by eliminating the geometrical factor $BL$,
\begin{equation}\label{WB}
 \frac{h}{m(\mathfrak{K})} = \frac{4gu}{n_1n_2n_3\nu^2} .
\end{equation}
All the quantities in the right-hand side of (\ref{WB}) can be measured with uncertainties small enough to give $h/m(\mathfrak{K})$ with relative uncertainty of less than $1\times 10^{-8}$, but, in practice, there are a number of uncertainties due, for example, to the alignments, unwanted motions, parasitic forces and torques.

\begin{figure}
\includegraphics[width=\columnwidth]{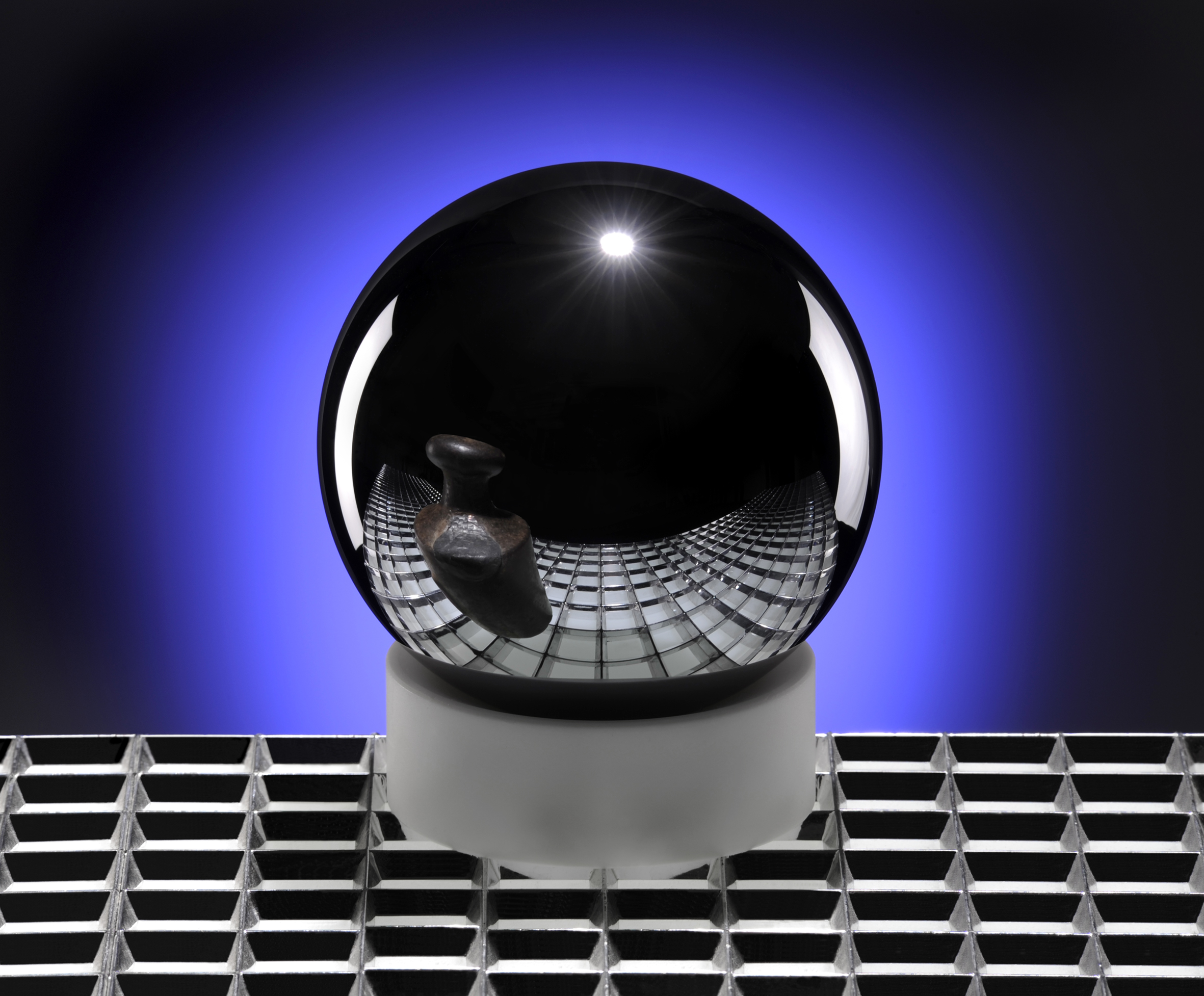}%
\caption{\label{sphere} Photograph of a $^{28}$Si sphere.}
\end{figure}

\section{The measurement of the Avogadro constant}
\subsection{Counting Si atoms}
A way to determine $h$ is by counting the atoms in 1 kg single-crystal spheres (see Fig.\ \ref{sphere}) that are highly enriched with the $^{28}$Si isotope. Today, this corresponds to determine the Avogadro constant. Since the relative atomic masses are extremely well measured, given measured values of $\NA$ and $h/m(\rmX)$, by rewriting
\begin{equation}
 \frac{h}{m(\rmX)} = \frac{\NA h}{M(\rmX)} ,
\end{equation}
where $M(\rmX)=A(\rmX) M_u$ is the molar mass, $A(\rmX)$ is the mass of the $\rmX$ isotope relative to $\rmC$, and $M_u=12$ g/mol is the molar mass constant, the Planck constant can be determined as well.
\subsection{Measurement equation}
Atoms are counted by exploiting their ordered arrangement in crystals; crystallization makes the lattice parameter accessible to macroscopic measurements, thus avoiding single atom counting. Provided the crystal and the unit cell volumes are measured and the number of atoms per unit cell is known, the counting requires their ratio to be calculated as
\begin{equation}\label{NA}
 \NA = \frac{nMV}{m a^3} ,
\end{equation}
where $n=8$ is the number of atom per cubic unit cell, $MV/m \approx 12.06$ cm$^3$/mol is the molar volume, $M \approx 27.977$ g/mol is the molar mass, $V \approx 431$ cm$^3$ and $m \approx 1$ kg are the crystal macroscopic volume and mass, $a^3 \approx 0.160$ nm$^3$ is the unit cell volume, and $a \approx 543$ pm is the lattice parameter. The measurement uses silicon crystals highly enriched with the $^{28}$Si isotope because, owing to the demands of modern electronics, they can be grown as high-purity, large, and quasi-perfect single crystals. Enrichment enables isotope dilution mass spectroscopy to be applied in determining the molar mass of the $^{28}$Si spheres with unprecedented accuracy, bypassing the limitations of the determination of the isotopic composition of natural silicon.

The following sections will outlines how the spheres' isotopic composition and chemical purity, molar mass, mass, volume, and lattice parameter were determined and their surfaces were geometrically, chemically, and physically characterized at the atomic scale.
\subsection{Molar mass}
Silicon occurs in three isotopes -- $^{28}$Si, $^{29}$Si, and $^{30}$Si. Therefore, a molar mass measurement to within a $10^{-8}M$ accuracy requires that the fractions of the minority isotopes $^{29}$Si and $^{30}$Si -- about 0.05 and 0.03, respectively -- are determined to within a relative accuracy better than $10^{-5}$. However, insuperable difficulties impaired the efforts to achieve this accuracy. To get around the problem the Central Design Bureau of Machine Building enriched a considerable amount of SiF$_4$ gas to more than 99.995\% $^{28}$SiF$_4$. Subsequently, the Institute of Chemistry of High-Purity Substances of the Russian Academy of Sciences converted the enriched gas into SiH$_4$ and grew a polycrystal. Eventually, a 5 kg $^{28}$Si crystal was grown and purified by the Leibniz-Institut f\"ur Kristallz\"uchtung. Since, in an enriched $^{28}$Si crystal, the minority isotopes contribute to the molar mass only through very small corrective terms, measurements of the minority-isotope fractions having a $10^{-2}$ relative uncertainty are sufficient \cite{Mana:2010u}. Accurate molar-mass measurements were thus possible by using a combination of isotope dilution mass spectrometry and high resolution inductively coupled plasma mass spectrometry \cite{Mana:2010m,Pramann:2011}.
\subsection{Volume}
Two 1 kg spheres were carved from the enriched $^{28}$Si crystal. The spherical shape was selected because it has no edges that might get damaged, because its volume can be calculated from diameter measurements, and because accurate geometric, chemical, and physical characterizations of the whole surface are possible. To weigh the spheres against the international prototype of the kilogram, their mass is 1 kg to within a few tens of a milligram. The sphere diameters -- 93.7 mm -- are measured by optical interferometry. In order to obtain a $10^{-8}$ relative accuracy in the volume determination, the mean diameter must be measured to a range of 0.3 nm, that is, to within an atom spacing. Such high accuracy requires sub-nanometre surface roughness and a quasi-perfect spherical shape. Eventually, the sphere volumes were determined to within a $2.9 \times 10^{-8}V$ uncertainty \cite{Bartl:2011,Kuramoto:2011}.
\subsection{Lattice parameter}
X-ray interferometry is the technology that enabled the measurement of the lattice parameter. An x-ray interferometer consists of three Si crystal slabs so cut that the \{220\} planes are orthogonal to the crystal surfaces. X-rays from a 17~keV~Mo~K$\alpha$ source are split by the first crystal and recombined, via two transmission crystals, by the third, called the analyser. When the analyser is moved in a direction orthogonal to the \{220\} planes, a periodic intensity-variation of the transmitted and diffracted x-rays is observed, the period being the diffracting-plane spacing. The analyser embeds front and rear mirrors, so that its displacement and rotations can be measured by an optical interferometer. The measurement equation is $\dd = m \lambda/(2n)$, where $\dd \approx 192$ pm is the spacing of the \{220\} planes, $n$ is the number of x-ray fringes in a displacement of $m$ optical fringes having period $\lambda/2 \approx 316$ nm, and the lattice parameter is obtained by $a=\sqrt{\dd}$. To ensure the interferometer calibration, the laser source operates in single-mode and its frequency is stabilized against that of a transition of the $^{127}$I$_2$ molecule. To eliminate the adverse influence of the refractive index of air and to achieve millikelvin temperature uniformity and stability, the experiment is carried out in a thermovacuum chamber. Continuous developments led a measurement accuracy of 3.5 nm/m \cite{Massa:2011}.
\subsection{Mass}
The BIPM, the PTB and the NMIJ carried out state-of-the-art comparisons between the masses of the $^{28}$Si spheres and Pt-Ir standards in air and under vacuum with a combined standard uncertainty of less than 5 $\mu$g. The mass of each sphere was determined by taking into account the traceability to the international prototype of the kilogram and the correlations among the 17 Pt-Ir standards used directly or indirectly in the comparisons \cite{Picard:2011}.

The surface of silicon is covered with a thin layer of silicon dioxide. The sphere surface was characterized from the chemical and physical viewpoints by X-ray reflectometry, X-ray photoelectron spectroscopy, X-ray fluorescence, near-edge X-ray absorption fine-structure spectroscopy, and spectroscopic ellipsometry to determine contamination, stoichiometry, mass, and thickness of the oxide layer with a spatial resolution of 1 mm$^2$ \cite{Busch:2011}.

The crystal must be free from imperfections and chemically pure. Consequently, it was purified by the float-zone technique and the pulling speed was so chosen in order to reduce the self-interstitial concentration. The crystal is dislocation free and, to apply the relevant corrections, the concentrations of carbon, oxygen, and boron atoms and vacancies were measured by infrared and positron life-time spectroscopies \cite{Zakel:2011}. Since, in determining the molar mass, consideration is given only to the Si atoms, the sphere masses were corrected for the mass of the surface layer and the bulk   point-defects, contaminants -- mainly carbon, oxygen, and boron -- and vacancies. In this way it was obtained the mass of an equivalent naked sphere having one Si atom at each lattice site.
\subsection{Results}
The $\NA$ values determined by using each of the two $^{28}$Si spheres differ only by $37(35)\times 10^{-9}\NA$. The average of these values, $\NA = 6.022 140 82(18)\times 10^{23}$ mol$^{-1}$, has a relative standard uncertainty of $3.0\times 10^{-8}$ and is the most accurate input datum for the determination of the Planck constant.

\section{Outlooks}
\begin{figure}
\includegraphics[width=\columnwidth]{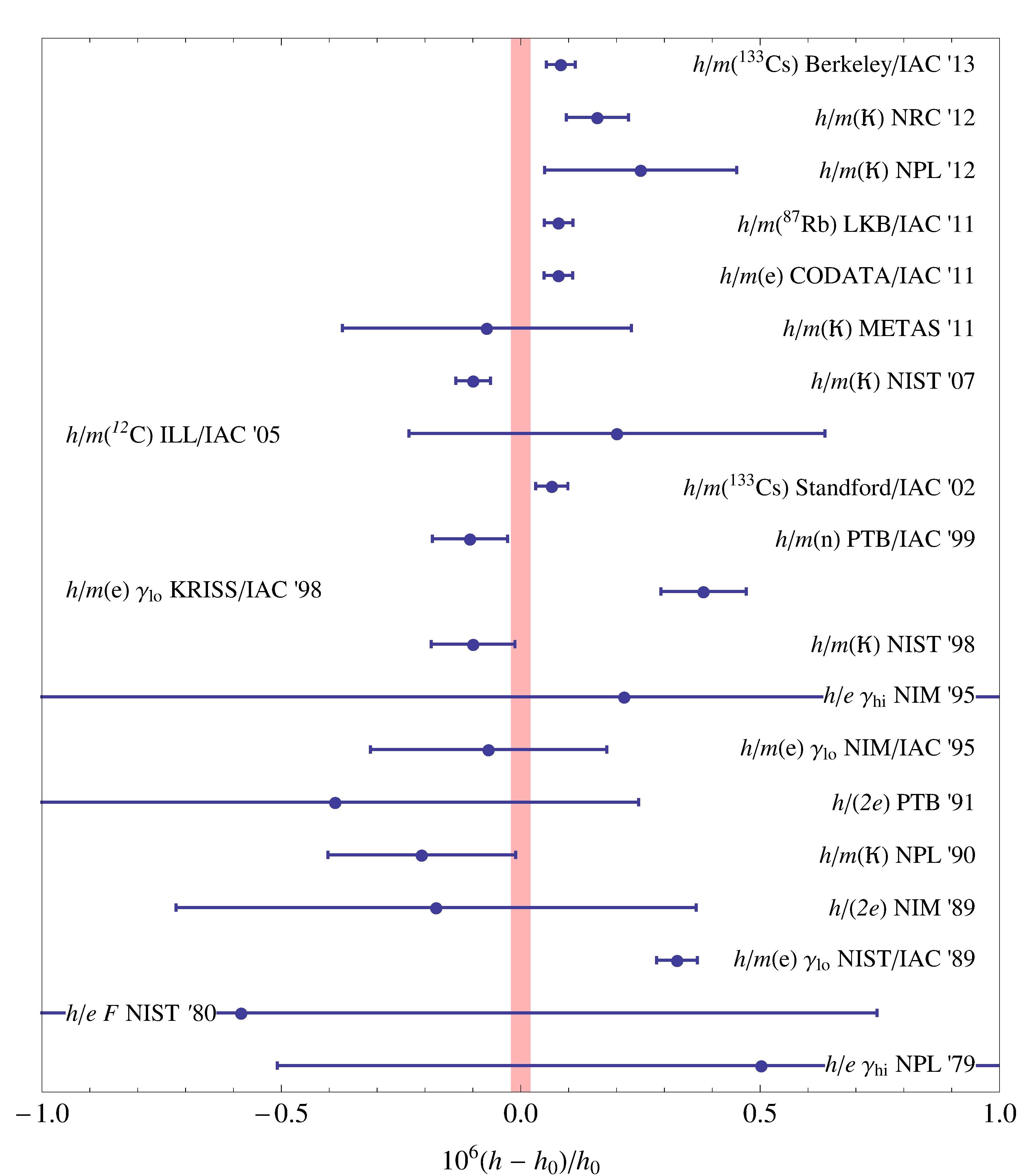}%
\caption{\label{Planck} Comparison of the determinations of the Planck constant -- data are form \cite{Mana:2012} and \cite{Mueller:2013}. The error bars indicate the standard deviations. The reference is the CODATA 2010 value, $h_0 = 6.62606957(29)\times 10^{-34}$ J s. The pink bar indicates the $\pm 2\times 10^{-8}h$ uncertainty required to make it possible a kilogram redefinition based on a conventionally agreed value of $h$.}
\end{figure}

Three $h$ measurements have so far achieved accuracies close to that required to make the kilogram redefinition possible: Two watt-balance experiments -- at the National Institute of Standards and Technology (NIST) \cite{Steiner:2005} and the National Research Council of Canada (NRC) \cite{Steele:2012} -- and a Si-atom count experiment carried out by the International Avogadro Coordination (IAC) \cite{Andreas:2011a,Andreas:2011b}. However, as shown in Fig.\ \ref{Planck}, there are inconsistencies between the values measured in these experiments: their spread, up to $2.6(7)\times 10^{-7}h$, is larger than the combined standard uncertainty. This indicates errors in at least one of the measurements; work is in progress to understand what this error is and to remedy it.

As regards the determination of $\NA$, the next paragraphs list the possible weakness and the activities necessary to investigate them experimentally. The long-term objective is to reach a $10^{-8}\NA$ measurement uncertainty; this stress test is expected to bring into light mistakes and hidden assumptions, excluding or identifying and eliminating them.

Temperature is a critical factor. Since the volumetric thermal expansion of Si is about $7.7 \times 10^{-9}$ mK$^{-1}$, the temperature measurements of the spheres and x-ray interferometer-crystals -- which are used for the volume and lattice parameter determinations -- require sub-mK accuracies. Luckily, measurements of thermodynamic temperature are not necessary, but it is required that the sphere and unit-cell volumes are referred to the same temperature. Therefore, it must be identified only the difference between the practical temperature scales used to extrapolate the volume and lattice-parameter measurements to 20 $^\circ$C. The thermal gradients in the relevant experimental set-ups must be investigated as well, both experimentally and by finite element modeling.

The total impurities within the $^{28}$Si crystal must not exceed few nanogram per grams; if higher, the relevant mass fractions must be quantified and the measured $\NA$ value must be corrected for. There is a general consensus that the contamination by the most of the elements are significantly smaller than $10^{-9}$ parts of a Si atom. In order to gain a direct evidence of purity, an analytical method based on neutron activation is being developed to exclude trace contaminations \cite{Dagostino:2012}.

The total-vacancy concentration in melt-grown Si crystals varies from zero -- if the crystal is grown in interstitial mode -- to a few times $10^{14}$ cm$^{-3}$; in general, there is a radial dependence of the vacancy concentration, too \cite{Falster:2010}. The emphasis on the total concentration is because, depending an a number of factors, vacancies react in various ways and become trapped in various forms during the cooling of the crystal -- primarily forming voids and/or complexes with oxygen and/or with nitrogen. It is probably safe to say that when the crystal has reached room temperature from solidification there are no free vacancies at all. The $^{28}$Si crystal used to determine $\NA$ was grown in vacancy mode; it is expected to contain roughly $3\times 10^{14}$ cm$^{-3}$ total vacancies \cite{Gebauer:1999,Martin:1999}. Since the number density of Si is $5\times 10^{22}$ cm$^{-3}$, in order to achieve a measurement uncertainty of $10^{-8}\NA$, the maximum total-vacancy concentration -- in any form or combination of forms -- must be below $5 \times 10^{14}$ cm$^{-3}$, if not there is a difficult problem.

A residual stress exists in silicon surfaces, also if the underlying bulk crystal is stress-free. When the surface relaxes, it strains the underlying crystal, makes the lattice parameters of the x-ray interferometer and spheres different, and jeopardize the atom count -- which is based on the atom spacing as measured in the interferometer crystals. Numerical techniques are being developed for first-principle characterization of both the stress and strain fields of silicon surfaces with atomistic resolution, as well as for the combination of surface stress with the continuum elasticity-theory. The experimental determination of the surface stress is a challenge; the design of a variable thickness interferometer is under way to work a lattice parameter measurement out so that there is a visible effect of the surface stress \cite{Quagliotti:2013}.

When the measurement accuracy approaches 1 nm/m, wavefront distortions are a major problem of dimensional metrology by optical interferometry. At this levels of uncertainty, the relation $\lambda=c/\nu$ (the symbols having the usual meanings) is valid only for a plane wave. In reality, some energy disperses outside the region in which it would be expected to remain in plane wave propagation. This effect is known as diffraction and is connected with the wave nature of light. As a result, wavefronts bend and their spacing varies from one point to another and is different from the wavelength of a plane wave. In the case of integrated signals, the analysis of the operation of two-beam interferometers proves that the relevant correction, in the limit of a small difference between the lengths of the paths through the interferometer arms, is proportional to the impulse-domain width of the illuminating beam, no matter how aberrated it could be \cite{Bergamin:1999}.

This result, provided that impulse-domain width is measured to within a sufficient accuracy, is central for the lattice parameter measurement, but it does not apply to wavefront waviness originated inside the interferometer or to position sensitive measurements. In this case the wavefront evolution with beam propagation makes the locally-detected phases different from those would be expected from a geometrical wavefront translation \cite{Andreas:2011}. This phenomenon plays a particular role in the reconstruction of the sphere topography. In fact, the roundness errors are not mapped one-to-one into the phase-profile ridges of the measured wavefront, but, in principle, they evolve from the sphere surface -- where they are imprinted on the wavefront -- to the detector -- where they are observed.

Whilst this evolution is believed small, a good estimate of the difference between the ridges of the phase profile at the detection plane and the roundness errors is still missing. Therefore, the Leibniz-Institut f\"ur Oberfl\"achenmodifizierung is developing technologies based on ultra-precision ion-beam figuring and plasma-jet machining to smooth the surface topography. A metallic contamination (by copper and nickel silicides) was detected on the sphere surfaces, which increased the measurement uncertainty because of an influence on the optical constants and mass of the oxide layer. The surface contamination, together with the roundness errors, was a major factor limiting the accuracy of the $\NA$ measurement.
\section{Conclusions}
The awaited revision of the SI assigns an exact value to the Planck constant; hence, the kilogram will be traced back to the second through the values of $h$, $c$, and $\nu(\Cs)$. Atomic masses are related to the Planck constant via the measurement of the $h/m(\rmX)$ quotient; the link to macroscopic masses is made by silicon spheres of known composition, volume, and lattice parameter. The number of atoms in such spheres is $N_{\rm Si} = 8V/a^3$ and, since the binding energy, about 5 eV per Si atom, is negligible, the sphere mass is $m_{\rm Si} = N_{\rm Si} M({\rm Si})/\NA$, where $M({\rm Si})$ is the molar mass and a correction must be applied for the mass of the surface oxide -- which must be characterized as regards the thickness and composition.

An immediate fallout of the $h$ and $\NA$ measurements is the possibility of monitoring the stability of the international prototype of the kilogram by using $^{28}$Si or natural Si spheres whose mass evolution is traced by monitoring the geometrical, physical, and chemical changes of their surfaces. These measurements play a role in science, too. They allow the consistency of our understanding of Nature to be investigated by checking the identity of the values measured at different energies, from meV (solid state physics) to eV (atomic physics and optical spectroscopy) to MeV (nuclear physics).

\section*{acknowledgements}
This work was jointly funded by the European Metrology Research Programme (EMRP) participating countries within the European Association of National Metrology Institutes (EURAMET) and the European Union.

\providecommand{\WileyBibTextsc}{}
\let\textsc\WileyBibTextsc
\providecommand{\othercit}{}
\providecommand{\jr}[1]{#1}
\providecommand{\etal}{~et~al.}

\end{document}